\documentclass[12pt]{iopart}
\usepackage{bm}
\usepackage{amssymb}
\usepackage{epsfig}
\usepackage{float}

\begin{document}

\def\rd{\mathrm{d}}

\title[Nondecaying modes in periodic dissipative lattice]{Nondecaying linear and nonlinear modes in a periodic array of spatially localized dissipations}

\author{S. C. Fern\'andez and V. S. Shchesnovich}

\address{Centro de Ci\^encias Naturais e Humanas, Universidade Federal do ABC, Santo Andr\'e, SP, 09210-170 Brazil}
%\ead{}
\begin{abstract}
 We demonstrate the  existence of  extremely weakly decaying linear and nonlinear  modes (i.e. modes immune to dissipation)  in the one-dimensional periodic   array of identical spatially localized dissipations, where the dissipation width is much smaller than the period of the array.  We consider   wave propagation governed by  the one-dimensional  Schr\"odinger equation in  the  array of identical Gaussian-shaped dissipations with   three parameters, the integral dissipation strength $\Gamma_0$,  the width $\sigma$ and the array period $d$. In the linear case, setting   $\sigma\to0$, while keeping $\Gamma_0$ fixed, we get an array of zero-width dissipations given by the Dirac delta-functions,  i.e. the  complex Kroning-Penney model,  where an   infinite number of nondecaying modes  appear with the  Bloch index being either at the center, $k= 0$, or at the boundary, $k= \pi/d $, of an analog of the Brillouin zone. By using numerical simulations we confirm that the weakly decaying modes persist for   $\sigma$ such that $\sigma/d\ll1$ and have the same Bloch index. The nondecaying modes persist also if  a  real-valued periodic potential is added to the spatially periodic array of dissipations, with  the  period  of the dissipative array  being  multiple of that of the periodic potential.  We also consider evolution  of the soliton-shaped pulses  in the nonlinear Schr\"odinger equation with the spatially periodic dissipative lattice and find that when the pulse width is much larger than the lattice period and  its wave number $k$  is either at the center, $k= 2\pi/d$, or at the boundary, $k= \pi/d $, a significant fraction of the pulse escapes the dissipation  forming    a stationary    nonlinear mode   with   the soliton shaped  envelope  and the Fourier spectrum consisting  of two peaks centered at  $k $ and $-k$.
\end{abstract}

 \maketitle

\section{Introduction}
\label{sec1}

The artificially structured materials such as photonic crystals have attracted  great interest for their remarkable properties to control and manipulate light. The propagation of radiation inside a photonic crystal is affected by the periodicity in a way similar to that of an electron traveling in a periodic potential in a solid state crystal, giving rise to the photonic analog of the band structures of solid-state physics ~\cite{Madelung}. An important starting point for this analysis is the Bloch theory which   provides the underlying mathematical framework for obtaining the fundamental wave propagation characteristics in photonic crystals. Through this theory, it is possible to obtain a relationship between the frequency and the wavenumber (wave vector). This relationship is referred as the frequency band structure.

On the other hand,  the absorption is likely to be an essential ingredient in future photonic structures, since purely dielectric arrangements have small index variation for opening  a complete bandgap below the infrared wavelengths. Indeed, there are proposals involving  the metallo-dielectric structures which are promising for   obtaining  the complete photonic bandgap \cite{MD1,MD2,MD3}. On the other hand, the absorptive periodic structures are not thoroughly investigated, with rather few results available. While a small  absorption can only slightly modify the band structure \cite{Krokhin},   in general,  the absorption turns Bloch bands into resonances in the lower-half  complex plane \cite{Band2Res,Moroz},  resulting in attenuation of the Bloch modes.  For instance, based on numerical simulations  of the periodic dielectric–-metallic superlattice,  it was concluded that the photonic band structure is strongly modified as compared to the band structure of the nonabsorbing superlattice: the Bloch vector becomes complex for all frequencies, and the waves are evanescent \cite{Soto}.   However, the influence of absorption seems to be far more complex than just  attenuation. Indeed, it has been known for quite some time that there are some interesting peculiarities in the effect  of action of dissipation on  the photonic band structures, such as the asymmetric behavior of the actual absorption rate   for wave vectors near the Brillouin-zone boundary and  the splitting of the lifetimes of the degenerate modes  \cite{Kuzmiak1,Kuzmiak2}.

Absorption attenuates the propagating modes but does this in a non-uniform way in the Fourier space of  the wave numbers.  Indeed, recently it was found that for a spatially periodic  localized dissipation, acting on the two-dimensional real-valued periodic lattice,  there are    some    modes featuring extremely weak dissipation, they appear  for the   Fourier index  lying   on the boundary of the Brillouin zone defined for the combined complex-valued lattice  \cite{EPL}. This important result was obtained for the two-dimensional   lattice which did not allow full understanding of the existence of  modes immune to the action of  dissipation.
It is of great interest to study this phenomenon in the  one-dimensional case,  which simplifies the analysis. Moreover,  the one-dimensional case allows for a rigorous introduction of a basis of dissipative modes due to the fundamental results in the Sturm-Liouville theory \cite{Marchenko}.

Here we consider the   one-dimensional linear and nonlinear Schr\"odinger equation with the complex-valued potential, concentrating on the results of the action of a periodic array of spatially  localized dissipations.  When the local dissipation width goes to zero (while the product of the amplitude and width remain fixed), our model  reduces to the complex  extension  of the well-known Kroning-Penney model (in the linear case),  which is a text-book illustration of the Bloch bands and Bloch waves \cite{Kronig-Penney1,Kronig-Penney2}.  The respective  eigenvalue problem, i.e. the  Sturm-Liouville problem with a complex-valued potential, is non-Hermitian. Fortunately, in the one-dimensional case there is  a fundamental result \cite{Marchenko}   showing the existence and completeness of the (generalized, in general) eigenfunctions. This  also  ensures the same property for the dissipative Bloch waves, as is discussed below.

As  was discussed before \cite{EPL}, besides application to the  propagation of light in photonic crystals,  our results  can have  applications in other branches of  physics as, for instance,  in the field of cold atoms and Bose-Einstein condensates in the spatially periodic traps created with laser beams, where   externally controlled     removal of  atoms serves as dissipation (see, for instance,   Refs. \cite{BKPO,S},  the removal of atoms can be done by a set of  ordered electron   beams with the use of the   electron microscopy technique of Refs. \cite{E1,E2}). In the Bose-Einstein condensates, the nondecaying modes illustrate also a very slow decay of the condensate subject to an external dissipation, i.e. the   quantum Zeno effect. The Zeno-like behavior in the Bose-Einstein condensates with a single dissipative defect was recently considered theoretically \cite{ZenoDefTh} and observed experimentally \cite{ZenoDefExp}.  Also a macroscopic quantum Zeno effect due to nonlinear interaction in the condensate was also recently studied \cite{ZenoSK}. For instance, our results suggest that such effects can be also  observed with  the spatially periodic dissipation.

The rest of the  text is organized as follows. In section \ref{sec2} the  general properties of the eigenvalue problem for a spatially periodic dissipative lattice are discussed. In section \ref{sec3} the imaginary Kronig-Penney  model is studied. Some  mathematical  details on the    dissipative Bloch waves, the respective Brillouin zone, and the completeness of the Bloch wave basis are relegated to the Appendix.  In section \ref{sec4} we numerically study the eigenmodes of the dissipative array of  localized finite-width  dissipations. In sections \ref{sec5}  we confirm the existence of the nondecaying modes by direct numerical solution of the   Schr\"odinger equation. In section \ref{sec6} we also consider the propagation of soliton pulses in the nonlinear Schr\"odinger equation with the dissipative lattice.  Summary of the results is given in section \ref{sec7}.

\section{Bloch waves and Bloch bands for a dissipative lattice}
\label{sec2}

We consider propagation  governed by the one-dimensional Schr\"odinger  equation and focus on the effect of a spatially  periodic localized  dissipation applied to  a periodic potential (lattice),
\begin{equation}
i\partial_{t}\psi = -\frac{1}{2}\partial_{x}^{2}\psi+\mathcal{V}(x)\psi \equiv \mathcal{L}\psi.
\label{E1}\end{equation}
It proves to be of great  convenience to introduce for a periodic dissipative lattice $\mathcal{V}(x)=V(x)-iG(x)$,  $\mathcal{V}(x+d)=\mathcal{V}(x)$, $V\in\mathbb{R}$ and $G\geq0$,  the Bloch waves $\psi_{n,k}(x)$ as   the eigenvectors of the non-Hermitian eigenvalue problem (i.e. for a complex-valued spatially periodic potential) with the standard  Bloch-like boundary condition:
\begin{equation}
\mathcal{L}\psi_{n,k}(x) = \mu_n(k) \psi_{n,k}(x), \; \psi_{n,k}(x+d) = e^{idk}\psi_{n,k}(x),
\label{E2}
\end{equation}
where $-\pi/d<k<\pi/d$ (an analog of the Brillouin zone). Here the reciprocal lattice basis is as usual and the eigenvalue problem is considered on  the unit cell of the dissipative lattice. The complex eigenvalues, $\mu_{n}(k)=E_{n}(k)-i\gamma_{n}(k)$ with $E_{n}(k)\in\mathbb{R}$ and $\gamma_{n}\geq0$, enumerated by a discrete  index $n$,  represent an analog of Bloch bands (now consisting, generally,  of  decaying modes).  Moreover, similarly to the real-valued potential case, there is a complete basis of such generalized Bloch waves  (a detailed treatment of the complex-valued potential and the generalized Bloch theory can be found in the Appendix).  Note also for the following  that for $k$ outside the Brillouin zone, the   eigenfunction  $\psi_k(x)$ is the same as that for an equivalent $k$ inside the zone. 

Our main interest lies in the spatially periodic dissipation which is sharply localized function in each period, thus  a particular profile of the dissipation near its maxima is not important.  We will use the Gaussian profile due to analytical simplicity, fixing its  period to be $2\pi$:
\begin{equation}
G(x) = G_0\sum_{m=-\infty}^\infty e^{-(x-2m\pi)^2/\sigma^2},
\label{E4}\end{equation}
where the narrow-width approximation requires that $\sigma \ll 2\pi$ (previously it was established that the nondecaying modes appear in this limit \cite{EPL}).

Our purpose is to find out properties of the nondecaying spatial structures  governed by  Eqs. (\ref{E1})-(\ref{E4}). Due to the spatial periodicity of the dissipation, it is sometimes simpler to work in the Fourier representation. We have:
\begin{equation}
G(x) = \sum_{n=-\infty}^\infty \Gamma_n e^{-inx}, \quad \Gamma_n = \frac{\sigma G_0}{2\sqrt{\pi}}e^{-(\frac{n\sigma}{2})^2}.
\label{E5}\end{equation}

By analogy with the real-valued periodic potential, one can rewrite the  dissipative Bloch-like wave as $\psi_k(x) = e^{ikx}u_k(x) $, $u_k(x+2\pi) = u_k(x)$, thus   the following expansion is valid
\begin{equation}
\psi_k(x) = e^{ikx}\sum_{n=-\infty}^\infty\,C_n(k) e^{-inx},
\label{E6} \end{equation}
where, in difference with the real-valued  case, the dissipative Bloch wave corresponds to a complex eigenvalue $\mu$. Assuming the solution $\psi_k(x,t) = e^{-i\mu t}\psi_k(x)$ with $\psi_k(x)$ of Eq. (\ref{E6}), we obtain the eigenvalue problem (\ref{E2})  in the form
\begin{equation}
\mu C_n = \frac12(k-n)^2C_n+\sum_{m=-\infty}^\infty(\hat{V}_{n-m}-i\Gamma_{n-m})C_m \equiv \sum_{m=-\infty}^\infty L_{n,m}(k)C_m,
\label{E7}\end{equation}
where $\hat{V}_n$ is the Fourier component of the real-valued potential $V(x)$.
Note that when $\sigma\ll 1$, about $1/\sigma$  terms in the Fourier series  (\ref{E5})  have  the amplitude $\Gamma_n \approx\Gamma_0$. This fact   allows for the extremely weakly decaying modes in Eq. (\ref{E7}), as we will see below, and the existence of such can be understood as a dissipative  analog of the well-known Bragg resonance.

\section{The zero-width dissipation approximation: the imaginary Kronig-Penney model}
\label{sec3}

For $\sigma\to 0$ with  $G_0\sim \sigma^{-1}$ (i.e. keeping $\Gamma_0 $ fixed) and $V(x)=0$ Eq. (\ref{E7}) reduces to the eigenvalue problem for the    Kronig-Penney model (see Ref. \cite{Kronig-Penney1}) with an imaginary potential, since  $\frac{1}{\sqrt{\pi}\sigma}e^{-\frac{x^2}{\sigma^2}} \to \delta(x)$ as $\sigma\to0$ and we obtain
\begin{equation}
G(x) \to 2\pi\Gamma_0\sum_{m=-\infty}^\infty \delta(x - 2m\pi).
\label{E18} \end{equation}
The eigenfunctions in this limit can be easily found using the Fourier representation (\ref{E7}). We have for the periodic part
\begin{equation}
u_{k}(x) =  -iu_{k}(0)\Gamma_0  \sum_{m=-\infty}^\infty \frac{e^{imx}}{\mu(k)-\frac12(m+k)^2},
\label{E19} \end{equation}
where the eigenvalues are obtained from the equation
\begin{equation}
\sum_{m=-\infty}^\infty\frac{1}{\mu(k)-\frac12(m+k)^2}= \frac{i}{\Gamma_0}.
\label{E20}\end{equation}
Eq. (\ref{E20})  is the consistency condition for Eq. (\ref{E19}) and can be cast as
\begin{equation}
\cos(2\pi k) = \cos( 2\pi\sqrt{2\mu}) -i \frac{2\pi\Gamma_0}{\sqrt{2\mu}}\sin(2\pi \sqrt{2\mu}),
\label{E21}\end{equation}
by noticing that the Mittag-Leffler expansion formula (i.e. the expansion in terms of the  pole singularities) for the following expression $\frac{\pi}{2\sqrt{\mu}}\left[\cot(\pi[\sqrt{\mu}+k])+ \cot(\pi[\sqrt{\mu}-k])\right]$ is the l.h.s. of Eq. (\ref{E20}). Eq. (\ref{E21}) is a complex-valued extension of that for  the ordinary  Kronig-Penney model (see, for instance,  Ref. \cite{Kronig-Penney2}).  It shows that the Kronig-Penney model with the imaginary potential (\ref{E18}) allows for the nondecaying modes when $\sin(2\pi \sqrt{2\mu}) = 0$, i.e. for $k = 0$ or $k = 1/2$ with $ \mu_{2n}(0) = \frac12n^2$ and  $\mu_{2n-1}(\frac12) = \frac12(n-\frac12)^2$, $n=1,2,3,\ldots\,$.

Let us now find the structure of the corresponding nondecaying modes. Since    they are eigenfunctions corresponding to the resonant points, i.e. to zeros of the denominator in Eq. (\ref{E19}),  and satisfy also $u_k(0)=0$ one has to deal with the undeterminate ratio of the type $0/0$ (i.e. the contributing  terms in the expansion of Eq. (\ref{E19}) are   those with zero denominator). To open the indeterminacy we recall the original equation for the eigenfunctions in the Fourier space, of Eq. (\ref{E7}), where for   $\Gamma_n = \Gamma_0$ under the condition that $\frac12(k-n)^2 = \mu(k)$ the sum on the r.h.s. must satisfy $\sum_{n=-\infty}^\infty C_n(k) = u_k(0)=0$. Hence,    the  nondecaying modes   are given by  a sum of just two    resonant Fourier terms with the amplitudes $C_{n_2}=-C_{n_1}$: for $k=0$, i.e. in the center of the Brillouin zone, we have  $n_2 = -n_1$ and  the nondecaying modes are  $u_0(x,n) = A_n\sin (nx)$, while for $k=1/2$, i.e. at the boundary  of the Brillouin zone, we have $n_2 = 1-n_1$ and the nondecaying modes are   $u_\frac{1}{2}(x,n) = B_ne^{-ix/2}\sin([n-1/2]x)$ (here   $n\equiv n_1\in\{1,2,3,\ldots\}$). We conclude noting that the real eigenvalues $\mu(k)$ \textit{belong either to the center or  the boundary of the Brillouin zone and this  resonant feature   can be understood as an analog of the Bragg resonance for the spatially periodic  dissipative lattice}.

\section{The  Bloch modes in the case of a periodic array of finite-width dissipations}
\label{sec4}

\begin{figure}[H]
%\vskip 1cm
\begin{center}
\epsfig{file=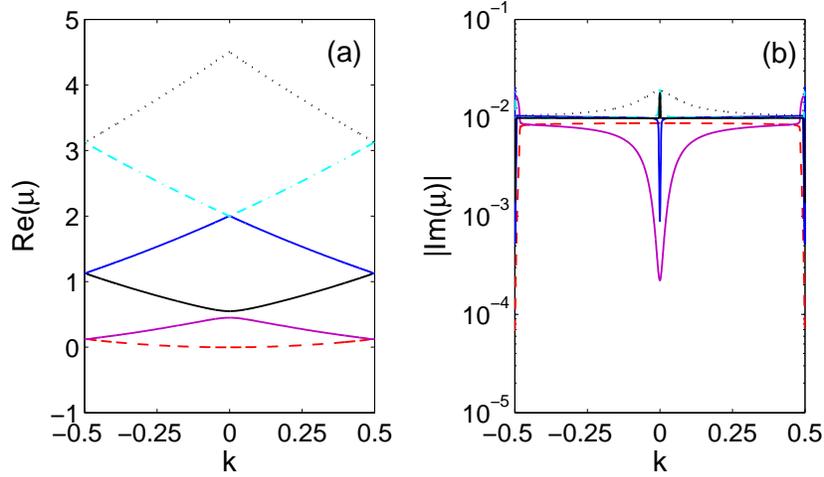,width=0.75\textwidth} \caption{  Panel (a), the real component  of the Bloch bands. Panel (b), the imaginary component of the Bloch bands.  Here $V_{0}=0.1$, $G_{0}=0.22$, $\sigma=\pi/20$, giving $\Gamma_{0}=0.01$. }
\label{F1}
\end{center}
\end{figure}

In the previous section we have analyzed the zero-width dissipation case and found an infinite number of nondecaying modes in the periodic  array of such dissipations.  Now let us consider the finite-width case. We concentrate on  the following periodic lattice  potential
\begin{equation}
\mathcal{V}(x)=V_0\cos(2x) -iG_0\sum_{m=-\infty}^\infty e^{-(x-2m\pi)^2/\sigma^2}.
\label{E23}\end{equation}
Properties of the Bloch band structure of this dissipative periodic lattice crucially   depend on the ratio of $\sigma$, the width of the dissipations,  to the distance between the sites, i.e.  the lattice period ($d=2\pi$ in our case), and on the dissipation strength $G_0$. Sometimes we will use the set of parameters $\sigma$ and  $\Gamma_0$  which are more appropriate  in the Fourier space. All numerical simulations of this and the following sections  are performed with the use of the highly accurate  Fourier  pseudospectral method (see, for instance, Ref. \cite{Pseudo}). Let us present the numerical results on the band structure of the dissipative lattice in Eq. (\ref{E23}). Fig. \ref{F1}  shows the real and imaginary parts of the Bloch bands for the lattice in Eq. (\ref{E23}).   Note the several magnitudes difference between the imaginary components. The  nondecaying modes appear only for $k=0$ or $k=1/2$. The  dissipation also partially closes the band gaps between the bands, only the   gap between the second and third bands is visible.

\begin{figure}[H]
%\vskip 1cm
\begin{center}
\epsfig{file=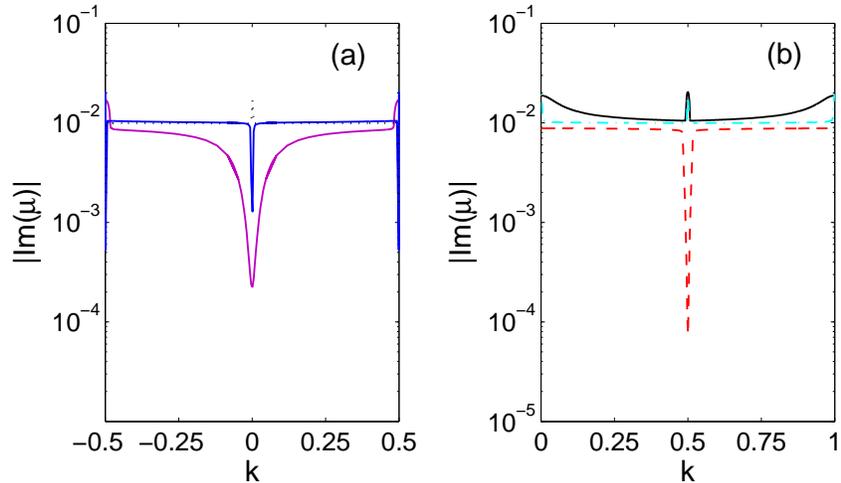,width=0.75\textwidth} \caption{ Panel (a), the imaginary component of the even bands. Panel  (b), the same for the  odd bands. The parameters are as in Fig. \ref{F1}. }
\label{F2}
\end{center}
\end{figure}

Fig. \ref{F2} shows  the imaginary component  of Fig. \ref{F1} separately for  the even and odd bands (enumerated by the real component). We can see that only  three bands remain  nondecaying at the center or at the boundary of the Brillouin zone (namely, the first, the second, and the fourth bands).  The even bands have a  nondecaying mode  in the center ($k=0$), whereas the odd band has a nondecaying mode at the boundary ($k=1/2$), i.e. in  agreement with the Kroning-Penney model,  though  a real-valued potential (with a compatible period) is added to the dissipation.

\begin{figure}[H]
\vskip 0.5cm
\begin{center}
\epsfig{file=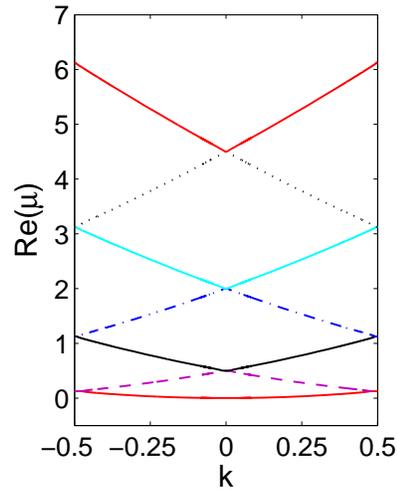,width=0.34\textwidth} \caption{  The real component of the Bloch bands for pure dissipative lattice.   Here $V_{0}=0$, $G_{0}=1.6$, $\sigma=\pi/100$, giving   $\Gamma_{0}=0.14$.}
\label{F3}
\end{center}
\end{figure}

\begin{figure}[H]
\vskip 0.5cm
\begin{center}
\epsfig{file=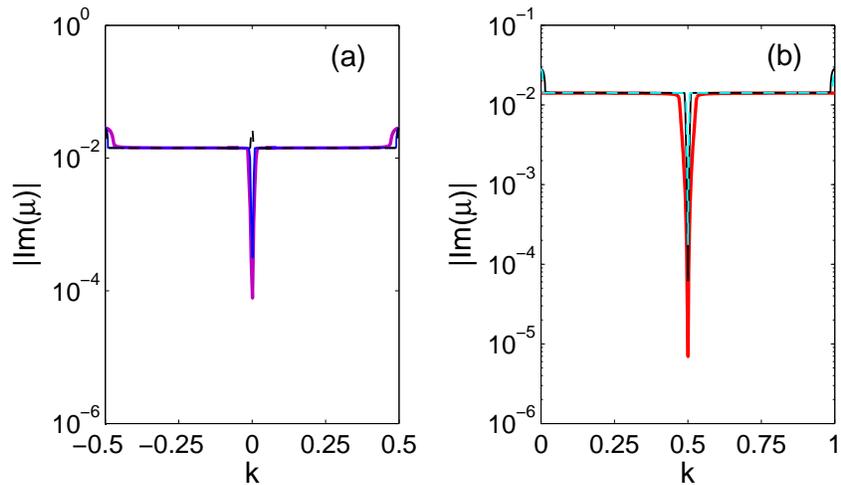,width=0.75\textwidth} \caption{ Panel (a), the  imaginary components of the even bands. Panel (b), the same for the  odd   bands. The  parameters are as in Fig. \ref{F3}.}
\label{F4}
\end{center}
\end{figure}

Thus, it becomes apparent that the main cause of the nondecaying modes in the realistic dissipative  lattice ($\sigma >0$) is the cancellation of the dissipation in the Fourier space, as it is in the imaginary Kronig-Penney model (however, the additional  real lattice term  can prevent  some of the cancellations, as it occurred with  the odd bands in Fig. \ref{F2}, where there is just one odd band with nondecaying modes as compared with Fig. \ref{F4}, where three odd bands have nondecaying modes). Therefore, for the rest of this section we consider the purely dissipative case, i.e. setting  $V(x)=0$. Fig.~\ref{F3}  confirms that for  the pure periodic dissipation  there are no band gaps (as in the imaginary Kronig-Penney model).  Fig. ~\ref{F4} shows   the imaginary components of the Bloch bands.   

\begin{figure}[H]
\vskip 0.5cm
\begin{center}
\epsfig{file=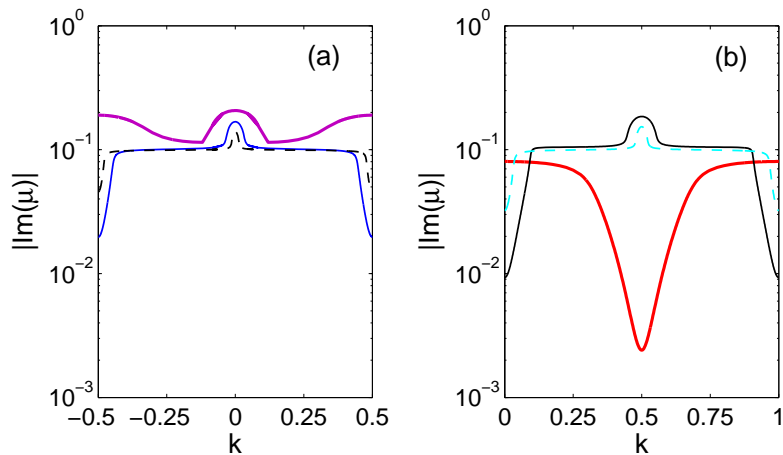,width=0.7\textwidth} \caption{   Panel (a) gives  the real part, panel (b) gives  the imaginary part. Here $V_{0}=0$, $G_{0}=1.12$, $\sigma=\pi/10$, giving a value of $\Gamma_{0}=0.1$.}
\label{F5}
\end{center}
\end{figure}

We note that the qualitative behavior of the Bloch  bands does not change with decrease of $\Gamma_{0}$, if   $\sigma$ is kept  constant. For very small values of $\sigma$, the  effects of the dissipation are the same as in the imaginary Kronig-Penney model with, however, finite dissipation  of the modes  in the center and at the boundary of the Brillouin zone (the dissipation  is stronger in the  higher bands).  Increasing   $\sigma$  results in    the  increase of the dissipation rate  in all bands, the weakly decaying modes disappear from the higher bands first.  All weakly decaying modes  disappear as $\sigma$ approaches the lattice period,   compare  Figs. \ref{F4} and \ref{F5}.  There is an additional feature for  $\sigma$ just an order smaller than the lattice period. For $\sigma=\pi/10$, as  in Fig. \ref{F5},  the higher bands have minima of the imaginary component in the switched places, i.e. the higher even bands   have minima at the boundary of the Brillouin zone, see panel (a) of  Fig. \ref{F5},  while the higher odd bands   have minima at the center,  panel (b) (note that the bands are labeled by their real component as in Fig. \ref{F3}).

\begin{figure}[H]
\vskip 0.5cm
\begin{center}
\epsfig{file=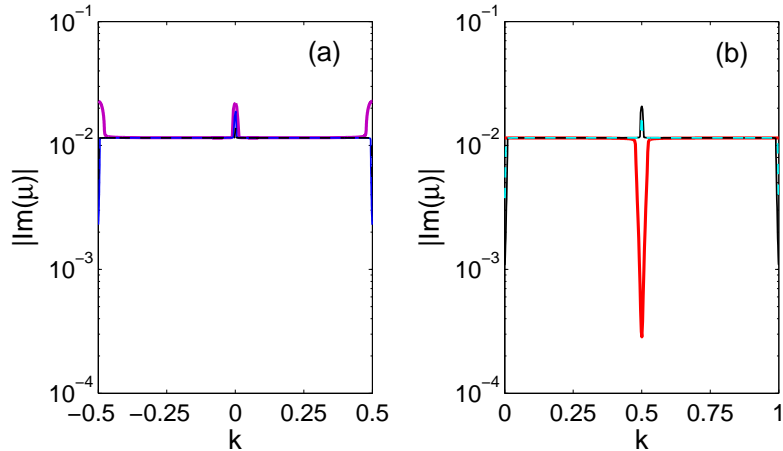,width=0.7\textwidth} \caption{Panel (a), the even imaginary bands.  Panel (b), the odd imaginary bands. Here $V_{0}=0$, $G_{0}=0.13$, $\sigma=\pi/10$, giving  $\Gamma_{0}=0.01$}
\label{F6}
\end{center}
\end{figure}

Now, by keeping the value of  $\sigma$ constant (we chose $\sigma=\pi/10$) we observe  that the effect  of decreasing   the parameter  $\Gamma_0$, the dissipation  strength in the Fourier space,  is to decrease  the width of  the minima of the  imaginary component of the Bloch bands, besides an obvious decrease of  their  values, compare Figs.~\ref{F5} and~\ref{F6}.

Therefore, we have established that  the appearance of extremely weakly decaying Bloch modes in a periodic dissipative array of finite-width localized dissipations  stems from the  cancellation of the dissipations in the Fourier space of  wave numbers, similarly as it is predicted by the imaginary Kronig-Penney model of section \ref{sec3} (i.e. we have a dissipative analog of the Bragg resonance). For a finite dissipation width $\sigma$, only a finite number of the nondecaying modes survive  (which  have   extremely  small dissipation rates, orders of magnitude smaller than the applied dissipation rate). We have found that the width of the dissipative analog of the Bragg resonance is  proportional to the strength of the dissipation in the Fourier space of wave numbers. Moreover, whereas a real-valued potential can modify  the Bloch bands and prevent for some of the cancellations of the dissipation (i.e. some of the dissipative Bragg resonances)   it has no  influence on the Bloch indices of the weakly decaying modes.  Below we consider the numerical  simulations of the time-dependent Schr\"odinger equation with  the dissipative lattice to confirm the predictions of the eigenvalue analysis.

\section{Wave dynamics    in the  dissipative  lattice}
\label{sec5}

\begin{figure}[H]
\begin{center}
\epsfig{file=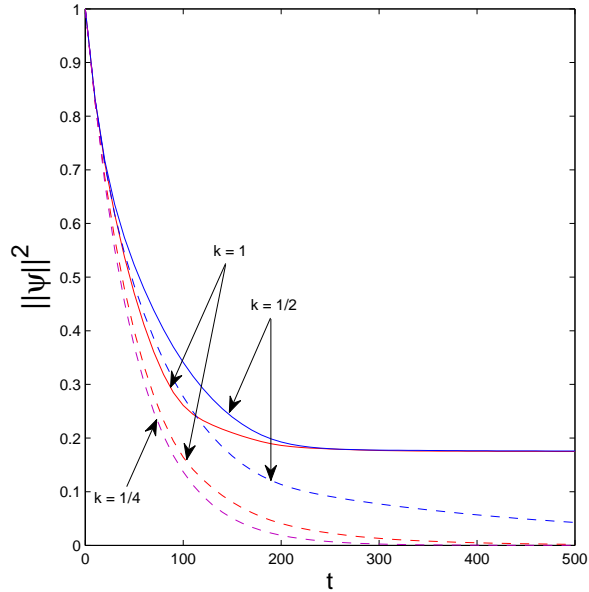,width=0.5\textwidth} \caption{ The norm of the solution  for  the initial condition $\Psi(x,0)=e^{ikx}U_{0}$. We use several Bloch indices: $k=1$, $k=1/2$, and $k=1/4$, for two values of $(\sigma,G_{0})$ (and one for $k=1/4$): $\sigma=\pi/100$ and $G_{0}= 1.13$ (solid lines) and $\sigma=\pi/4$ and $G_{0}= 0.045$ (dashed lines). Here $V_{0}=0$ and $\Gamma_{0}= 0.01$.}
\label{F7}
\end{center}
\end{figure}

For  numerical  solution of the Schr\"odinger equation we use  the so-called split-step Fourier method \cite{Pseudo} with the  pseudo-spectral  approximation of  the derivatives in the spatial coordinate. According to the discussion above, we  focus on  the   purely dissipative lattice, setting  $V_{0}=0$.    Without any loss of generality, we can use a plane wave $\psi(x,0)=e^{ikx}U_{0}$ as the initial condition. Since the dissipative lattice $\mathcal{V}$ is periodic, evolution governed by Eq. (\ref{E1}) preserves the quasimomentum $k$ and   assumes the form $\psi(x,t)=e^{ikx}U(x,t)$ (see also the Appendix). We can thus simulate the time evolution on the primitive lattice cell (simulations made  on a finite lattice for  the Gaussian-shaped initial condition of  width much larger than the lattice period confirm this). The results are given in Fig. \ref{F7}, where we plot the rescaled norm of the solution, i.e. $\|\psi\|^{2}=\int\psi(x,t)|^{2}dx/\int|\psi(x,0)|^{2}dx$ (where  the integration is  over the lattice period $2\pi$). One can clearly spot  the   presence of extremely weakly decaying modes. Moreover, they exist only for the wave index lying either in the center $(k=1)$ or at the boundary $(k= 1/2)$ of the  Brillouin zone (of the dissipative lattice). Here we note that Bloch modes with    the wave number    shifted by  addition of  multiples of the Fourier period $K=\frac{2\pi}{d}=1$   belong to  higher Bloch bands, thus $k\pm n$,  with $k$ in the  Brillouin zone,  may be nondecaying for some natural $n$ depending on the value of $\sigma$ (depending on the value of  $\sigma$ there are  several  bands with the weakly decaying modes, as is discussed in the previous section).

\begin{figure}[H]
\begin{center}
\epsfig{file=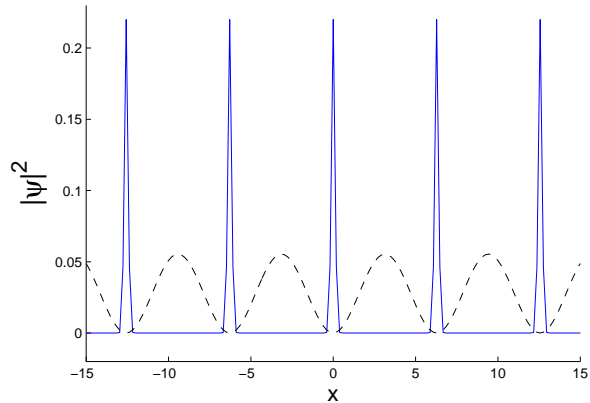,width=0.5\textwidth} \caption{ The  spatial profile of the nondecaying wave solution.     We show the dissipative sites (the solid  line) and the   solution (the dashed line).   The parameters correspond to one curve  from  Fig. \ref{F7}, namely, $k=1/2$,  $\sigma=\pi/100$, $G_0=1.13$, and  $\Gamma_0=0.01$.}
\label{F8}
\end{center}
\end{figure}

As  the nondecaying Bloch mode   must approach  zero at the maxima of the dissipation, i.e. in  intervals of size on the order of  $\sigma$, the nondecaying waves  of Fig. \ref{F7}   have periodic variation of the absolute value with period of the dissipative array,  see Fig. \ref{F8}. They are linear combination of the weakly decaying Bloch modes present in  the lower bands.

%%%%%%%%%%%%%%%%%%%%

\section{Nonlinear wave dynamics in dissipative lattice}
\label{sec6}

Consider now  propagation of nonlinear waves governed by the  nonlinear Schr\"odinger equation  with the periodic dissipative lattice, i.e.
\begin{equation}
i\partial_{t}\Psi = -\frac{1}{2}\partial_{x}^{2}\Psi+\mathcal{V}(x)\Psi -  g|\Psi|^2\Psi.
\label{EQ1} \end{equation}
Setting $V=0$ (i.e. considering  the purely dissipative lattice) and $g>0$ (to allow for the soliton solutions for zero dissipation), let us  focus on the effect of a spatially periodic dissipation on the nonlinear wave dynamics.

Our model can be considered as a finite-width extension of an  imaginary variant of the nonlinear Kroning-Penney model.  The existence of  localized (soliton-like) and extended (Bloch-like) nonlinear modes   in the usual nonlinear  Kroning-Penney  model (i.e. with a real-valued potential)   was studied before (see for instance, Refs. \cite{NonlKP1,NonlKP2} and references therein), where exact solutions were found.    Here we also note that in the case when $V(x)$ is  a real-valued periodic potential (not necessarily localized within each period)  Eq. (\ref{EQ1})  possesses   the so-called gap soliton solutions, due to the fact that they  appear in the band gaps of the  Bloch spectrum of the periodic  potential.  The gap solitons   were  first  discovered  in  optics  \cite{OptGapSol1,OptGapSol2}  (see also a book \cite{OptSolBook}) and then  predicted \cite{BookGapSol} and observed experimentally in the quasi one-dimensional Bose-Einstein condensates \cite{GapSolBEC}. In contrast, the periodic dissipation  does not lead to the band gaps, thus our nonlinear  waves, which are discussed below,  are quite different from both the gap solitons and the nonlinear Bloch modes studied before.

\begin{figure}[H]
%\vskip 1cm
\begin{center}
\epsfig{file=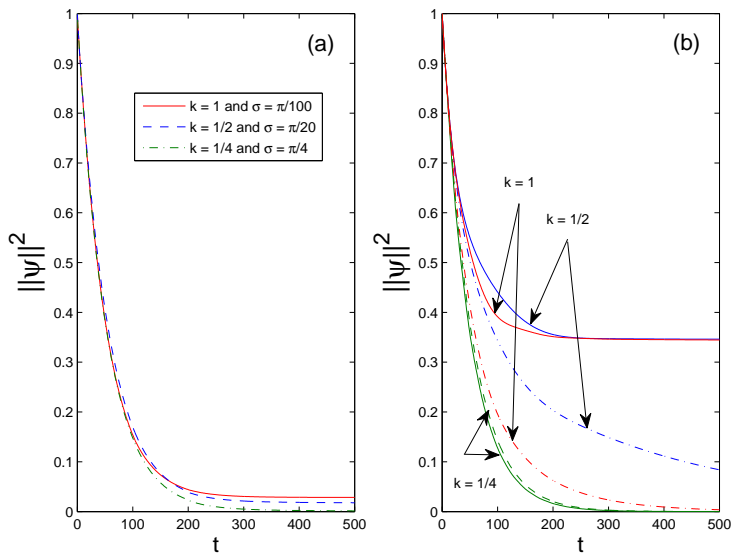,width=0.65\textwidth} \caption{ Panel (a), the norm of the solution for $g=1$ and the initial condition $\Psi(x,0)=e^{ikx}A\sqrt{\frac{2}{g}}$\rm{sech}$(Ax)$ with $A = 1/\pi$. Panel (b), the norm of the solution when $A = 1/(10\pi)$. We use several Bloch indices: $k=1$, $k=1/2$, and  $k=1/4$, for three values of $(\sigma,G_{0})$: $\sigma=\pi/100$ and $G_{0}= 1.13$ (the solid lines), $\sigma=\pi/20$ and $G_{0}= 0.22$ (the dashed lines), and $\sigma=\pi/4$ and $G_{0}= 0.045$ (the dash-dotted lines). In all cases $\Gamma_0 = 0.01$. }
\label{F9}
\end{center}
\end{figure}

We find that there are nonlinear extremely  weakly decaying  modes,  i.e. nonlinear  waves immune to the applied periodic dissipation, see Fig. \ref{F9}.  As the  initial condition we  have used   the usual soliton profile $\Psi(x,0)=e^{ikx}A\sqrt{\frac{2}{g}}$\rm{sech}$(Ax)$, of the nonlinear Schr\"{o}dinger equation (note that   $A=1/l$, where $l$ is the width of the soliton).  When $l \sim 2\pi$ all waves decay, see Fig. \ref{F9}(a), whereas for $l\gg2\pi$  the waves   with the wave index  in the center of the Brillouin zone of the dissipative lattice (we use $k=1$)   and at its boundary  ($k=1/2$) are almost immune to the dissipation, see Fig. \ref{F9}(b).

\begin{figure}[H]
%\vskip 1cm
\begin{center}
\epsfig{file=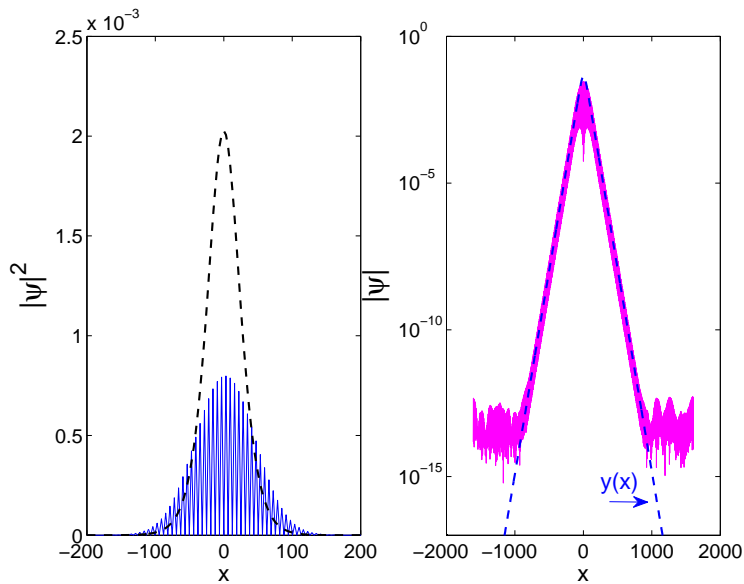,width=0.65\textwidth} \caption{ Panel (a),  the nonlinear (soliton) wave solution   for the initial time (the  dashed line) and at a  much larger time (the solid line).    Panel (b), the envelope of the solution at a large time    (the solid line) compared with  $y(x)=A_1\sqrt{\frac{2}{g}}\mathrm{sech}(A_1 x)$ with $A_1 = \mathrm{max}(|\psi(x,500)|)$ (the dashed line).   Here   $k=1/2$, $A = 1/(10\pi)$, $\sigma=\pi/100$, and $G_{0}=1.13$.}
\label{F10}
\end{center}
\end{figure}

The nondecaying nonlinear waves in the dissipative lattice can be considered as   analogs of the soliton solutions  of  the usual (i.e. conservative)  nonlinear  Schr\"odinger equation.     Indeed,  in Fig.~\ref{F10} we  show the spatial structure of such a  nonlinear  nondecaying  wave in the dissipative lattice.  Fig.~\ref{F10}(b) shows that  the spatial envelope of the nondecaying nonlinear  wave  has  the soliton shape.   In Fig.~\ref{F10} we use $k=1/2$, but for   $k=1$ the result is very similar (in both cases there are extremely weakly decaying  nonlinear waves in the model, as shown in Fig.~\ref{F9}(b)). Moreover, to verify that the nonlinear wave of Fig.~\ref{F10} contains the nonlinear eigenvalue of the Riemann-Hilbert problem associated with the canonical nonlinear Schr\"{o}dinger equation, and thus indeed contains a soliton, we have numerically verified the respective sufficient condition \cite{SolitonBook}, which is formulated as follows
\begin{equation}
I (t)= \int\limits_{-\infty}^\infty dx\, |\psi(x,t)| \ge \mathrm{ln}(2+\sqrt{3}).
\label{I}\end{equation}
Condition  (\ref{I}) must be satisfied for large  values of $t$,  if the soliton part  is present in the  nonlinear wave. Here we note that condition (\ref{I}) is formulated only for the nonlinear Schr\"odinger  equation without the dissipative lattice, however, one can use the nonlinear spectrum provided by the Riemann-Hilbert problem to judge about the presence of a soliton component in the solution for the perturbed equation,  similarly as it was done in Ref.~\cite{PhysD} for an exact treatment of the soliton-radiation interaction in a nonintegrable evolution  equation of the nonlinear Schr\"{o}dinger type.  We have verified that  in the case of  the nonlinear wave of Fig.~\ref{F10},  and in all other cases, when the localized wave survives the dissipation, the  behavior of $I(t)$ is as follows: it  initially rapidly decreases  (for instance, in the case of Fig.~\ref{F10} on an interval $t\sim 10$) to a constant value well satisfying condition (\ref{I}) and then stays almost constant. 

\begin{figure}[H]
%\vskip 1cm
\begin{center}
\epsfig{file=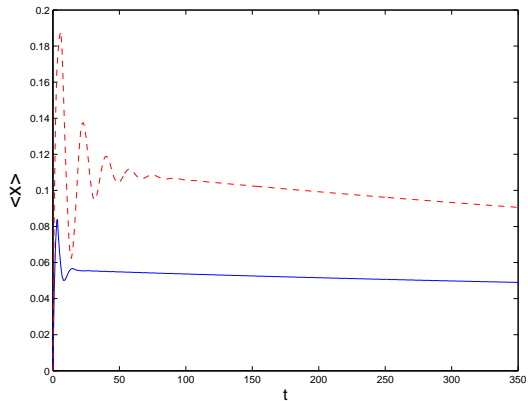,width=0.45\textwidth} \caption{ The  spatial position of the nonlinear wave. Here $k=1/2$ (the solid line) and $k = 1$ (the dashed line). The other parameters are as in Fig. \ref{F9}. }
\label{F11}
\end{center}
\end{figure}

Finally, we  note that the soliton waves  with the wave number $k$ lying at the center or on the boundary of the Brillouin zone (for instance,  we used  $k=1/2$ and   $k=1$) is  attracted to a \textit{stationary} (extremely weakly decaying)  nonlinear mode.  This is illustrated in Fig. \ref{F11}, where we plot the time evolution of the pulse mean  position
\begin{equation}
\langle x \rangle  = \frac{\int dx x |\Psi|^2}{\int dx |\Psi|^2}.
\end{equation}
One can clearly see the signatures that the solution first acquires some   negative Fourier components (which results  in the spatial oscillations of the mean position) and then settles at a stationary position in the dissipative lattice.  Moreover, we have verified that the Fourier spectrum of the stationary nonlinear mode    is indeed  composed of $k$ and $-k$ peaks of equal amplitude, as one would expect from the above analysis due to  periodicity of the array of dissipations. Thus,  the stationary mode can be rightfully called a bound soliton,  due to its soliton-shaped  envelope  and the symmetric Fourier spectrum about the origin.

\section{Conclusion}
\label{sec7}
We have found the existence of extremely weakly decaying linear and nonlinear waves in the one-dimensional spatially periodic array of localized dissipations, when the dissipation width is much smaller than the array period. The origin of such nondecaying  modes was established by considering the zero-width limit of the model, which is a complex-valued generalization of the well-known Kronig-Penney model of the solid state physics. The Kronig-Penney model with an imaginary potential allows for an infinite number of nondecaying Bloch modes with the wave index lying either at the center or at the boundary of the corresponding Brillouin zone (a generalization of the usual concept for the complex-valued potential). We have numerically confirmed that the nondecaying (i.e. almost nondecaying in this case) modes persist in the periodic  array of  finite-width dissipations, however, only  a few  lowest Bloch bands allow for such  modes for   the dissipations of a finite width.  The lower-band nondecaying modes survive  if a real-valued potential is added to the dissipative lattice, when the period of the dissipative array  is multiple of that of the real-valued potential.  Numerical simulations of the governing   Schr\"odinger equation confirm the existence of the  linear  waves immune to the dissipation. Moreover, we have studied the nonlinear  Schr\"odinger equation, which allows for the soliton solutions in the absence of the dissipation, and found that a significant portion of the soliton pulse survives the dissipative lattice forming  a novel stationary nondecaying nonlinear wave.  Moreover,   the  envelope of the emerging nonlinear stationary wave is of the soliton shape, it has the symmetric Fourier spectrum with respect to the index inversion and thus can be rightfully  called the soliton bound state of the model, which is immune to the dissipation.  The results can have immediate applications in the propagation of optical pulses in the photonic crystals and in the field of cold atoms and Bose-Einstein condensates in the periodic optical lattices.

\section{Acknowledgements}  VSS was supported by the CNPq  and  SCF was supported by the CAPES of Brazil.

\appendix
\section{ The  Bloch wave theory generalized to the spatially  periodic dissipative lattice}

Similar as is in the case of the usual Bloch waves in  a real-valued potential, the dissipative Bloch waves can be used for expansion of an arbitrary function $f(x)$, decaying for large $x$ at least as $ \sim |x|^{-3}$, if we  allow for the generalized Bloch-wave solutions to  the eigenvalue problem, $\psi^{(s)}_{k,n}(x)$, labeled by an index $s$ for each eigenvalue. They  satisfy
\begin{equation}
\mathcal{L}\psi^{(s)}_{k,n}(x) = \mu_n(k) \psi^{(s)}_{k,n}(x) + \psi^{(s-1)}_{k,n}(x),  \quad \psi^{(s)}_{n,k}(x+d) = e^{idk}\psi^{(s)}_{n,k}(x),
\label{E8} \end{equation}
where $\mathcal{L}=-\frac{1}{2}\partial_{x}^{2}+\mathcal{V}(x)$ with a complex-valued periodic potential $\mathcal{V}(x)$,  $\psi^{(0)}_{k,n}(x) \equiv \psi_{k,n}(x)$ is the  analog of the usual Bloch wave (i.e. the eigenmode  to which a Bloch wave transforms when a periodic dissipation of infinitesimal amplitude is added to the real-valued lattice potential). Indeed, assuming the period of the  combined complex-valued  lattice to be $d= 2\pi$,  the Fourier representation of $f(x)$ can be transformed as follows
\begin{equation}
f(x) = \int\limits_{-\infty}^\infty\rd k\, e^{ikx} \hat{f}(k) =  \int\limits_{-\frac12}^\frac12 \rd k\, e^{ikx}\hat{F}(k,x),
\label{E9} \end{equation}
where we have identified $\hat{F}(k,x) = \sum_{n=-\infty}^\infty e^{inx}\hat{f}(k+nk)$, which is periodic function in $x$: $\hat{F}(k,x+2\pi) = \hat{F}(k,x)$. By a well-known result on the generalized Sturm-Liouville problem for a complex-valued periodic potential \cite{Marchenko}, the spatially periodic function $\hat{F}(k,x)$ can be expanded over the   basis of the generalized eigenvectors of the non-Hermitian Sturm-Liouville problem, formulated for $\mathcal{L}$  on the interval $x\in [0,2\pi]$ (moreover, the multiplicity $m_n$ of each complex eigenvalue $\mu_{n}(k)$ is finite, i.e. $0\le s\le m_n$ in Eq. (\ref{E8})). That is to say that $e^{ikx}\hat{F}(k,x)$ can be expanded over the basis of the generalized Bloch waves $\psi^{(s)}_{k,n}(x)$ with a fixed $k$, where the index $n$ enumerates the eigenvalues $\mu_n(k)$. Therefore, we get the expansion for an arbitrary $f(x)$:
\begin{equation}
f(x) = \int\limits_{-\frac12}^\frac12 \rd k\,  \sum_{n=1}^\infty\sum_{s=0}^{m_n}b_{n,s}(k)\psi^{(s)}_{k,n}(x).
\label{E10} \end{equation}
The coefficients $b_{n,s}(k)$ can be obtained by using the inner product with the eigenvectors of the adjoint (i.e.  Hermitian conjugated)  eigenvalue problem, formulated  for $\mathcal{L}^\dag=\mathcal{L}^*$:
\begin{equation}
\mathcal{L}^\dag \tilde{\psi}^{(s)}_{k,n}(x) = \mu^*_n(-k)\tilde{\psi}^{(s)}_{k,n}(x) + \tilde{\psi}^{(s-1)}_{k,n}(x)
\label{E11}\end{equation}
with the same Bloch-type boundary condition. The adjoint eigenvectors are related to the eigenvectors of $\mathcal{L}$ by
\begin{equation}
\tilde{\psi}^{(s)}_{k,n}(x) = (\psi^{(s)}_{-k,n}(x))^*.
\label {E12}
\end{equation}
Eq. (\ref{E11}) simplifies for a potential satisfying the inversion property $\mathcal{V}(-x) = \mathcal{V}(x+a)$ (satisfied by our dissipative lattice). Indeed, we have in this case $\mu_n(-k) = \mu_n(k)$. The inner product relations can be established by a standard procedure with the use of Eqs. (\ref{E8}) and (\ref{E11}), they read
\begin{eqnarray}
\int\limits_{-\infty}^\infty\rd x\, (\tilde{\psi}^{(\tau)}_{k',n'}(x))^*\psi^{(s)}_{k,n}(x) &&= \int\limits_{-\infty}^\infty\rd x\, \tilde{\psi}^{(\tau)}_{-k',n'}(x)\psi^{(s)}_{k,n}(x) \nonumber \\
&&=\delta(k-k')\delta_{n',n}\delta_{\tau,m_n-s}.\qquad
\label{E13}\end{eqnarray}

For example,  a  solution to the linear Schr\"odinger equation $i\partial_tU(x,t) = \mathcal{L}U(x,t)$ can be expanded over the dissipative Bloch waves as follows
\begin{equation}
U(x,t) = \int\limits_{-\frac12}^\frac12 \rd k\,  \sum_{n=1}^\infty e^{-i\mu_n(k)t}\sum_{s=0}^{m_n}\frac{(-it)^s}{s!}b_{n,s}(k)\psi^{(s)}_{k,n}(x),
\label{E14} \end{equation}
where
\begin{equation}
b_{n,s}(k) = \int\limits_{-\infty}^\infty\rd x\, \psi^{(m_n-s)}_{-k,n}(x)U(x,0).
\label{E15}\end{equation}
Note that only weakly or  nondecaying Bloch waves are important for the long-time behavior of solutions to the Schr\"odinger equation. Thus one does no need to characterize all the generalized eigenvectors, restricting oneself only to the weakly decaying ones.

In difference with the Hermitian eigenvalue problem, generally, the  generalized eigenvectors are common,  they are solutions to Eq. (\ref{E8}) (i.e. the  multiplicity $m_n>0$ for some $\mu_n(k)$). However, their appearance is conditioned on that
\begin{equation}
\int\limits_{0}^{2\pi}\rd x\, u_{-k,n}(x)u_{k,n}(x)=0,
\label{E16} \end{equation}
which follows from Eq. (\ref{E13}).

Now let us return to the specific eigenvalue problem (\ref{E7}) of  section \ref{sec2}. First of all, we note that the operator $\mathcal{L}$ (\ref{E1}) is symmetric with respect to inversion $x\to-x$, resulting in $L_{n,m}(k) = L_{m,n}(k)$, thus $u_{-k,n}(x) = u_{k,n}(-x)$, $C_n(-k) = C_{-n}(k)$. Hence, the orthogonality for the proper eigenvectors in Eq. (\ref{E13}) simplifies to
\begin{equation}
\sum_{n=-\infty}^\infty C'_n(k) C_n(k) = 0, \quad \mu'(k)\ne\mu(k),
\label{E17}\end{equation}
and a similar sum appears in  condition (\ref{E16}) for the  generalized eigenvectors. Condition (\ref{E16})    was checked numerically in our case,  we have found no generalized eigenvectors  for  the eigenvalues $\mu(k)$ which correspond to the  weakly decaying modes, i.e. the mode of our principal interest.

\end{document}